\documentclass[12pt]{article}
\usepackage{latexsym}
\usepackage{amsmath,,calrsfs}
\usepackage{amsfonts}
\usepackage{amssymb}
\usepackage{amscd}
\usepackage{bbm}
\usepackage{fancybox}
\usepackage{cite}
\usepackage{amsmath,amsfonts,amsbsy}
\usepackage{pstricks,pst-node}
\usepackage[small,bf,hang]{caption2}
\usepackage{graphicx,xcolor}
\usepackage{epsfig}
\usepackage{psfrag}
\usepackage{comment}

\usepackage{float}

\psset{unit=1.3cm,linewidth=.5pt,radius=.2}  

\usepackage{multirow}                     
\usepackage{float}                          
\usepackage{lscape}                         
\usepackage{bm}

\newcommand{\beq}{\begin{equation}}
\newcommand{\eeq}{\end{equation}}
\newcommand{\bt}{\begin{tabular}}
\newcommand{\et}{\end{tabular}}
\newcommand{\bc}{\begin{center}}
\newcommand{\ec}{\end{center}}

\newcommand{\be}{\begin{equation}}
\newcommand{\ee}{\end{equation}}
\newcommand{\bea}{\begin{eqnarray}}
\newcommand{\eea}{\end{eqnarray}}
\newcommand{\ba}{\begin{array}}
\newcommand{\ea}{\end{array}}

\def\bbox{{\,\lower0.9pt\vbox{\hrule \hbox{\vrule height 0.2 cm
\hskip 0.2 cm \vrule height 0.2 cm}\hrule}\,}}
\newcommand{\dsl}{\pa \kern-0.5em /}

\makeatletter \@addtoreset{equation}{section} \makeatother

\def\slashchar#1{\setbox0=\hbox{$#1$}           
   \dimen0=\wd0                                 
   \setbox1=\hbox{/} \dimen1=\wd1               
   \ifdim\dimen0>\dimen1                        
      \rlap{\hbox to \dimen0{\hfil/\hfil}}      
      #1                                        
   \else                                        
      \rlap{\hbox to \dimen1{\hfil$#1$\hfil}}   
      /                                         
   \fi}

\date{}

\begin{document}

\begin{titlepage}

\begin{center}

\vskip 1.5cm

{\Large \bf  Born Again}
\smallskip

\vskip 1cm

{\bf Jorge G.~Russo\,${}^{a,b}$ and  Paul K.~Townsend\,${}^c$} \\

\vskip 25pt

{\em $^a$  \hskip -.1truecm
\em Instituci\'o Catalana de Recerca i Estudis Avan\c{c}ats (ICREA),\\
Pg. Lluis Companys, 23, 08010 Barcelona, Spain.
 \vskip 5pt }

\vskip .4truecm

{\em $^b$  \hskip -.1truecm
\em Departament de F\' \i sica Cu\' antica i Astrof\'\i sica and Institut de Ci\`encies del Cosmos,\\ 
Universitat de Barcelona, Mart\'i Franqu\`es, 1, 08028
Barcelona, Spain.
 \vskip 5pt }
 
 \vskip .4truecm

{\em $^c$ \hskip -.1truecm
\em  Department of Applied Mathematics and Theoretical Physics,\\ Centre for Mathematical Sciences, University of Cambridge,\\
Wilberforce Road, Cambridge, CB3 0WA, U.K.\vskip 5pt }

\hskip 1cm

\noindent {\it e-mail:}  {\texttt  jorge.russo@icrea.cat, pkt10@cam.ac.uk}

\end{center}

\vskip 0.5cm
\begin{center} {\bf ABSTRACT}\\[3ex]
\end{center}

Born's original 1933 theory of nonlinear electrodynamics  (in contrast to the 
later Born-Infeld theory) is acausal for strong fields. We explore the issue of
strong-field causality violation in families of theories containing Born and/or
Born-Infeld, and many variants that have been previously proposed in contexts that
include cosmology and black hole physics. Many of these variants are acausal 
and hence unphysical.  A notable exception is the modified Born-Infeld theory 
with ModMax  as its conformal weak-field limit.

\vfill

\end{titlepage}
\tableofcontents

\section{Introduction}

In 1933 Born introduced the first Lorentz-invariant (and gauge-invariant) nonlinear theory of electrodynamics (NLED) \cite{Born:1933qff}. He was motivated by the idea of an electromagnetic origin for the electron mass, which he supposed might be finite in a nonlinear extension of Maxwell electrodynamics if the electric field had some maximum value. An analogy with the maximum velocity of relativistic particle mechanics led him to propose a Lagrangian density of the form\footnote{We take the opposite overall sign from Born, and add a constant to get zero vacuum energy.} 
\be\label{LBorn}
\mathcal{L}_{\rm Born} = - \sqrt{T^2 -2TS} +T\, , \qquad S = \frac12\left(|{\bf E}|^2 - |{\bf B}|^2\right), 
\ee
where $T$ is a positive constant with dimensions of energy density\footnote{$T$ is related to Born's parameter $a$ (which  later became $b$) by $Ta^2=1$.},
and $S$ is the quadratic Lorentz scalar expressed in terms of the electric and magnetic\footnote{Born uses the notation 
${\bf H}$ for magnetic field in \cite{Born:1933qff}, which is confusing because of the standard definition 
${\bf H} = -\partial\mathcal{L}/\partial{\bf B}$, but ${\bf H}$ was replaced by ${\bf B}$ in \cite{Born:1933lls}.}
fields $({\bf E},{\bf B})$. In the weak-field limit (equivalent to $T\to\infty$)  $\mathcal{L}_{\rm Born} \to S$, which is the Maxwell Lagrangian density in 
appropriate units, but reality of the Born Lagrangian density requires the inequality 
\be
|{\bf E}|^2 \le T+ |{\bf B}|^2\, , 
\ee
which puts an upper bound on $|{\bf E}|$ for any given ${\bf B}$. 

Later in 1933 \cite{Born:1933pep}, and again in more detail in 1934 \cite{Born:1934gh}, Born and Infeld  jointly proposed 
a modified Lagrangian density; in our notation this is
\be\label{BI}
\mathcal{L}_{\rm BI} = - \sqrt{T^2 -2TS -P^2}+T\, , \qquad P= {\bf E}\cdot {\bf B}\, , 
\ee
where $P$ is the quadratic Lorentz pseudoscalar.  This was also given in the equivalent manifestly Lorentz invariant form
\be\label{BI.det}
\mathcal{L}_{\rm BI} = - T\sqrt{-\det\left(\eta + F/\sqrt{T}\right)} +T\, , 
\ee
where $\eta$ is the standard Minkowski-space metric (with ``mostly-plus'' signature) and $F$ is the antisymmetric 
matrix of components of the 2-form field-strength $F=dA$ for 1-form potential  $A= dt A_0+ {\bf dx}\cdot {\bf A}$. 
In this form the BI theory has a natural generalisation to higher dimensional Minkowski spacetimes, although this was not
a consideration at the time. The BI alternative was initially presented as an illustration of the point that an electromagnetic origin for the electron mass did not, by itself, determine the required nonlinearities. Born explored some of the properties of the BI extension of his theory in a 1937 review article  \cite{Born:1937drv}. In particular, he discusses the ``self-duality under 
Legendre transform'' of the BI theory, which is now understood (see e.g. \cite{Kuzenko:2000uh}) to be a consequence of its $U(1)$ electromagnetic invariance, a property first noticed by
Schr\"odinger \cite{Schrodinger:1935oqa} that was also reviewed by Born. 

Neither Born's original theory nor its BI extension is now seen as particularly relevant to electrodynamics, classical or quantum; in that domain, the 1936 Euler-Heisenberg low-energy effective theory of QED is much more significant \cite{Heisenberg:1936nmg} (see \cite{Dunne:2012vv} for a relatively recent review). However, the special features of the BI theory continued to attract the attention of theorists (e.g. Dirac in 1960 \cite{Dirac:1960}) and a surprising new feature of the theory was found around 1970: BI is the unique NLED with a weak-field limit for which interactions do {\sl not} lead to birefringence (polarisation dependent dispersion relations) \cite{Boillat:1966eyw,Boillat:1970gw,Plebanski:1970zz}.  This result was originally found in the context of shock-wave propagation but can also be found by considering a constant uniform electromagnetic background as a (typically birefringent) homogeneous optical medium through which plane wave perturbations propagate; this reformulation was introduced by Bialynicki-Birula in a 1983 review of BI theory that also introduced the conformal strong-field limit of BI \cite{Bialynicki-Birula:1984daz}, which we refer to here as BB electrodynamics.  

In 1985 Fradkin and Tseytlin  showed how the BI theory in the form \eqref{BI.det}, but in a 26-dimensional Minkowski spacetime, arises as an effective low-energy field theory of open-strings in the bosonic open-string theory \cite{Fradkin:1985qd}. In 1989, this result was extended by Leigh to the effective low-energy description of D-branes \cite{Leigh:1989jq}; in this context the BI theory is extended to a Dirac-Born-Infeld (DBI) theory, which becomes a supersymmetric DBI theory for the D-branes of the 10-dimensional Type II superstring theories. In particular, 
the original BI theory in a 4-dimensional spacetime becomes relevant to the D3-brane of IIB superstring theory. There have since been many research papers on BI theory in the context of string/M-theory (see e.g. the 2000 review by Tseytlin \cite{Tseytlin:1999dj} and the 2018 
review by Gibbons \cite{Gibbons:2001gy}).

In comparison to this activity in the development and application of Born-Infeld theory, Born's original theory has been largely ignored. However, it has re-emerged in various contexts over the last 30-plus years, often as a special case of some class of nonlinear electrodynamics defined by a Lagrangian density that, like Born's theory, is a function of the Lorentz scalar $S$ but not of the 
pseudo-scalar $P$. Such ``Born-type'' theories, with Lagrangian density $\mathcal{L}(S)$, were applied to inflationary cosmology around 1990 \cite{Altshuler:1990} and more recently to black hole physics \cite{Ayon-Beato:1998hmi,Bronnikov:2000vy,Maeda:2008ha,Uniyal:2022vdu,Tomizawa:2023vir}. Other recent examples are an ``inverse electrodynamics'' proposed as a potential competitor to the Euler-Heisenberg theory in the event of a discovery of birefringence effects in laser physics \cite{Gaete:2021ytm}, and three families of Born-type theories applied to holographic superconductors \cite{Lai:2022tjr}. In light of this activity and the current revival of interest in NLED (see e.g. \cite{Sorokin:2021tge}), Born's original theory still warrants attention.

The initial impetus for this paper was a recent incidental observation that the strong-field limit of Born's original theory, which can be viewed as a $T\to0$ limit, differs from that of BI \cite{Mezincescu:2023zny}. This fact is not obvious from \eqref{LBorn} because a simple $T\to0$ limit of it cannot be taken. However, the $T\to0$ limit can be taken in a Hamiltonian formulation, and then a calculation of dispersion relations using the general ``Hamiltonian birefringence'' results of \cite{Mezincescu:2023zny} is very simple; it shows that there is always a range of directions in which one polarisation mode is propagated faster than light. This implies that the Born theory itself must be unphysical for sufficiently strong fields, in contrast to BI. It appears that the acausality of Born's theory was first noticed in a 2016 paper by Schellstede et al. \cite{Schellstede:2016zue} in which necessary and sufficient conditions for causal propagation were found for the ``Pleba{\'n}ski class'' of NLED defined (in our notation) by a Lagrangian density function $\mathcal{L}(S,P)$. 

Assuming $\mathcal{L}_S>0$, which is standard for reasons detailed in \cite{Schellstede:2016zue}, these conditions are 
\be\label{Schell1}
\mathcal{L}_{SS}\ge 0\, , \qquad \mathcal{L}_{PP} \ge 0\, , \qquad 
\mathcal{L}_{SS}\mathcal{L}_{PP} - \mathcal{L}_{SP}^2 \ge0 \, , 
\ee 
and 
\be\label{Schell2}
\mathcal{L}_S + S(\mathcal{L}_{SS} - \mathcal{L}_{PP}) +2P\mathcal{L}_{SP} > (\mathcal{L}_{SS} + \mathcal{L}_{PP})\sqrt{S^2+P^2}\, .  
\ee
The conditions \eqref{Schell1} are precisely those shown in \cite{Bandos:2021rqy} to be equivalent to convexity of $\mathcal{L}$ as a function of ${\bf E}$, which is required to avoid superluminal propagation in weak-field backgrounds \cite{Russo:2022qvz}. It also ensures the existence of an equivalent Hamiltonian formulation.  However, convexity is insufficient (generically) to exclude superluminal propagation in strong-field backgrounds. For this we need \eqref{Schell2}, for which we will provide an alternative derivation by consideration of causality bounds on birefringence indices.

Despite the relative complexity of the condition \eqref{Schell2}, it has a simple corollary for any NLED of the ``Pleba{\'n}ski class'' for which $\mathcal{L}$ is independent of $P$ (e.g. $\mathcal{L}_{\rm Born}$)  \cite{Schellstede:2016zue}. In this case \eqref{Schell2} reduces to
\be\label{F+G}
\mathcal{L}_S + S\mathcal{L}_{SS} > \mathcal{L}_{SS} \sqrt{S^2+P^2}  \qquad (\mathcal{L}_P\equiv 0)\, ,  
\ee
but this inequality is necessarily violated for sufficiently large $P$. This is because the left-hand side is independent of $P$ while the right-hand side is positive for an interacting theory of the specified type satisfying \eqref{Schell1},  and increases linearly with $P$ for large $P$. Notice that this argument depends on the fact that there is no upper bound on $P$. In contrast, reality of $\mathcal{L}_{BI}$ imposes an upper bound on $P$ for any given $S$, and in this case the inequality \eqref{Schell2} is satisfied, as expected from the earlier result of \cite{Bialynicki-Birula:1984daz} that BI does not allow superluminal propagation. 

This simple argument from  \cite{Schellstede:2016zue} rules out as unphysical not only Born's original theory but also all ``Born-type'' theories with Lagrangian density $\mathcal{L}(S)$, such as those mentioned above.  A variant  of it can be used to rule out many other NLED theories defined by a Lagrangian density of the form $\mathcal{L}=F(S)+G(P)$, where $(F,G)$ are functions only of $(S,P)$, respectively; the convexity/causality conditions are satisfied if 
$(F',F'',G'')$ are all positive. In such cases \eqref{Schell2} reduces to
\be\label{ffggg}
F'(S) > F''(S) (\sqrt{S^2+P^2}-S)+ G''(P)(\sqrt{S^2+P^2}+S)\, . 
\ee
Both sides are positive but the left-hand side is independent of $P$. The inequality is therefore violated for sufficiently large $P$ at any fixed $S$, unless reality of $G(P)$ imposes a sufficiently severe upper bound on $P$, which does not happen for polynomial $G$ (for example), and this observation rules out as unphysical several more NLED theories considered in the literature ({\it e.g.} \cite{Novello:1998ws,Gibbons:2001sx,Kruglov:2014iwa}).

For more generic theories in the Pleba{\'n}ski class the causality condition \eqref{Schell2} is still a significant constraint.
An example considered in  \cite{Schellstede:2016zue}
is the Heisenberg-Euler Lagrangian expanded to quadratic order in $(S,P)$. While the
weak-field causality conditions are satisfied, the strong field
causality condition is violated. This approximate theory is therefore acausal but for fields that are too strong for the validity of the approximation \cite{Schellstede:2016zue}. Whether the full Heisenberg-Euler theory is causal is unknown. 

For the remainder of this paper we investigate precisely how causality is violated (or not) in strong-field backgrounds, 
in the context of  models previously considered in the literature, and variants of them, that satisfy the simple
weak-field convexity/causality constraints. Our aim is to develop some further intuition into strong-field causality violation. 

A particularly instructive example is a natural one-parameter family including both Born and Born-Infeld 
that was introduced by Kruglov  \cite{Kruglov:2009he}; in our notation the Lagrangian density is
\be\label{Lxi}
\mathcal{L}^{(\xi)} =  T- \sqrt{T^2 -2TS -\xi^2 P^2}\, , 
\ee
where $\xi$ is a dimensionless parameter, which we may assume to be positive without loss of generality. We could replace  $\xi^2$ by an arbitrary real parameter but if this parameter is negative then the Lagrangian density is not convex and the causality conditions \eqref{Schell1} are violated. We therefore lose nothing by the parametrisation of \eqref{Lxi} and this one-parameter family includes the original Born theory ($\xi=0$) and the BI theory ($\xi=1$). As we shall see, the strong-field limit for any non-zero $\xi$ is the same as the Born-Infeld theory, i.e. the {\sl causal} BB theory. This led us to expect that the acausality of Born's original theory would disappear for $\xi\ne0$, but this does not happen. It turns out that $\xi=1$ is required for causality, which makes Born-Infeld the exceptional member of the family. It is also the only one that is electromagnetic duality invariant, and one may wonder how significant this is.

There are few duality invariant NLED theories for which the Lagrangian density function 
$\mathcal{L}(S,P)$ is known {\sl explicitly}. One is a deformation of BI that has the interacting conformal ModMax electrodynamics as its weak-field limit \cite{Bandos:2020jsw}; here we call it ModMaxBorn. The ModMaxBorn Lagrangian density depends on the BI parameter $T$ and a dimensionless coupling constant $\gamma$, and it reduces to BI for $\gamma=0$ \cite{Bandos:2020hgy}. For $\gamma<0$ the convexity/causality conditions \eqref{Schell1} are violated but they are satisfied for $\gamma>0$. We find here that the strong-field causality condition \eqref{Schell2} is also satisfied, thus establishing ModMaxBorn as a physical deformation of BI. 

It should be obvious that electromagnetic duality invariance does not guarantee causality 
(a simple counterexample is ModMax with $\gamma<0$ \cite{Bandos:2020jsw}). Neither is it true that  
electromagnetic duality invariance is required for causality; we provide counterexamples here.
Nevertheless, it appears to us that almost all NLED theories appearing in the literature that are not duality invariant are also not causal. We have put many to the test; most pass the convexity/causality test \eqref{Schell1} but few pass the strong-field causality test \eqref{Schell2}. The examples that we present here illustrate this state of affairs.

\section{Causality in strong-field backgrounds}\label{sec:causality}

For any NLED of the ``Pleba{\'n}ski'' class, the field equations are solved by 
any constant uniform $({\bf E}, {\bf B})$. Small-amplitude disturbances of this background propagate as electromagnetic waves with two independent polarisations, as in the vacuum,  but 
the dispersion relation for these waves is generically polarisation dependent in the homogeneous optical medium provided by the background; this is the phenomenon of birefringence. There are therefore two, generically distinct, dispersion relations for the wave 4-vector $(\omega,{\bf k})$, which may be characterised by a pair of birefringence indices $\lambda_\pm$. They take the form \cite{Bialynicki-Birula:1984daz,Russo:2022qvz}
\be\label{DR}
(1+ \lambda |{\bf E}|^2) \omega^2 + 2\lambda ({\bf k}\cdot {\bf S}) \omega 
= (1+ \lambda |{\bf E}|^2)|{\bf k}|^2 - \lambda |{\bf k} \times {\bf E}|^2 - \lambda |{\bf k} \times {\bf B}|^2\, , 
\ee
where ${\bf S}={\bf E}\times {\bf B}$, and $\lambda$ may be either $\lambda_+$ or $\lambda_-$. Convexity of $\mathcal{L}(S,P)$ as a function of ${\bf E}$ ensures that $\lambda_\pm\ge0$; equality for both $\lambda_+$ and $\lambda_-$ occurs only in the vacuum, for which the dispersion relation degenerates to the standard relativistic relation $\omega^2= |{\bf k}|^2$. 

For a generic constant uniform background, the vector field ${\bf S}$ is also constant and uniform but not necessarily zero, which means that the homogeneous medium in which electromagnetic waves propagate is generically stationary rather that static. However, generic stationary backgrounds are Lorentz boosts of static backgrounds and we may then choose the 
rest-frame of the medium, in which it is static; i.e. ${\bf S}={\bf 0}$. In this frame the 
dispersion relation \eqref{DR} simplifies to 
\be
\omega^2 = A |{\bf k}_\perp|^2 + k_\parallel^2\, , 
\ee
where ${\bf k}_\perp$ is orthogonal to the common direction of ${\bf E}$ and ${\bf B}$ ($k_\parallel$ is the remaining parallel component)  and 
\be\label{A}
A= \left(\frac{1-\lambda B^2}{1+\lambda E^2}\right)\, , 
\ee
where $E= |{\bf E}|$ and $B= |{\bf B}|$. As $\lambda$ is a Lorentz scalar, we may rewrite $A$ in a manifestly Lorentz invariant form by using the fact that in the background rest-frame $P=\pm EB$,  and hence 
\be\label{restframe}
E^2 = \left[\sqrt{S^2+P^2} + S\right]_{{\bf S}={\bf 0}}\, , \qquad 
B^2 = \left[\sqrt{S^2+P^2} -S\right]_{{\bf S}={\bf 0}} \, . 
\ee

The phase velocity is 
\be
v_{\rm ph} = \sqrt{\frac{A |{\bf k}_\perp|^2 + k_\parallel^2 }{
|{\bf k}_\perp|^2 + k_\parallel^2}}
\ee
This exceeds the velocity of light in vacuum only if $A>1$. However, it is generally the case that causality requires the group velocity $v_g = |d\omega/d{\bf k}|$ to be subluminal or lightlike, and this is \cite{Bialynicki-Birula:1984daz,Russo:2022qvz}
\be\label{vg}
v_g = \sqrt{\frac{A^2 |{\bf k}_\perp|^2 + 
k_\parallel^2}{A |{\bf k}_\perp|^2 + k_\parallel^2}}\, . 
\ee
Notice that $v_{\rm ph} \ge v_g$ with equality when $k_\parallel=0$.

Inspection of the formula \eqref{vg} shows that superluminal propagation is generic whenever $A>1$, and possible for certain choices of the wave-vector ${\bf k}$ if $A<0$. The necessary and sufficient 
conditions for causality are therefore the inequalities
\be\label{A-causal}
A \le 1 \, , \qquad A\ge0\, . 
\ee
The group velocity does not always coincide with the signal velocity\footnote{We thank Wenqi Ke for raising this point.};
 see \cite{Fox:1970cu} for a discussion. However, the two fail to coincide only for 
 absorptive or gainful materials, which  is not the case here. 
A relevant point in this context is that the light-cones defined by the dispersion relations used here were originally found by consideration of propagating shock-wave discontinuities  \cite{Boillat:1966eyw,Boillat:1970gw,Plebanski:1970zz}, which are clearly  signal fronts.

From \eqref{A} and \eqref{A-causal} we see that the  causality inequalities to be satisfied by $\lambda$ are 
\be\label{lambda-causal}
\lambda \ge 0 \, , \qquad \lambda B^2 \le 1\, . 
\ee
We also see that $\lambda<0$ allows $A>1$ even for weak fields, so we may view 
$\lambda\ge0$ (equivalently, $A\le1$) as a weak-field causality condition; it is equivalent to the conditions \eqref{Schell1} which, as mentioned in the Introduction, are also convexity conditions. 
In contrast, $A< 0$ is generically possible only for strong fields\footnote{An exception is conformal theories for which there is no weak/strong distinction for field strengths.}, so we may view $\lambda B^2\le1$ (equivalently $A\ge0$) as a strong-field causality 
condition. 

The main aim of this section is show how $A>0$ is equivalent to the causality condition \eqref{Schell2}, at least for propagation in static backgrounds. We exclude $A=0$ here, despite the fact that it is compatible with causality, because it restricts the direction of
wave-propagation, and this implies particular properties of the Hamiltonian density that are never satisfied by any NLED of the  ``Pleba\'nski'' class  \cite{Russo:2022qvz,Mezincescu:2023zny}. It is convenient to first rewrite \eqref{Schell2} as
\be\label{Schell2'}
\mathcal{L}_S \ > \ (\sqrt{S^2+P^2} + S) \mathcal{L}_{PP} +   (\sqrt{S^2+P^2} - S) \mathcal{L}_{SS} - 2P\mathcal{L}_{SP} \, . 
\ee
In the rest-frame of a static background we may use \eqref{restframe} to reduce this inequality to 
\be\label{Schell2reduced}
\mathcal{L}_S > E^2 \mathcal{L}_{PP} + B^2 \mathcal{L}_{SS} + 2\sigma EB \mathcal{L}_{SP}\, , 
\ee
where $P$ has been written as $P=\sigma EB$, where $\sigma=+1 (-1)$ for (anti)parallel ${\bf E}$ and ${\bf B}$. 

 To make contact with the causality conditions \eqref{A-causal} we need to know the birefringence indices for a generic NLED. They are determined by the first and second derivatives of $\mathcal{L}(S,P)$. Following \cite{Bialynicki-Birula:1984daz} (but using the slightly different notation of \cite{Russo:2022qvz}) we introduce 
the `normalized' second derivatives
\be
\ell_{SS} = \frac{\mathcal{L}_{SS}}{\mathcal{L}_S}\, , \qquad 
\ell_{SP} = \frac{\mathcal{L}_{SP}}{\mathcal{L}_S}\, , \qquad 
\ell_{PP} = \frac{\mathcal{L}_{PP}}{\mathcal{L}_S}\, , 
\ee
and the definitions
\be
\Xi := \frac12 \left(\ell_{SS}+ \ell_{PP}\right) \, , \qquad 
\Gamma := \ell_{SS}\ell_{PP} - \ell_{SP}^2 \, ,  
\ee
and
\be\label{J}
J := 1 - P^2\Gamma + 2(P\ell_{SP} - S\ell_{PP}) \, .   
\ee
The convexity/causality conditions \eqref{Schell1} are now 
\be\label{Schell1'}
\ell_{SS}\ge0\, , \qquad \ell_{PP} \ge0 \, , \qquad \Gamma \ge0\, . 
\ee 
Notice that the first two conditions can be replaced by $\Xi\ge0$ when $\Gamma>0$.
We may now write the birefringence indices as
\be\label{biref.indices}
\lambda_\pm = J^{-1}\left[(\Xi-S\Gamma) \pm \sqrt{(\Xi-S\Gamma)^2-\Gamma J} \right]\, . 
\ee
This presupposes that $J\ne0$. The existence of a weak-field limit implies $J>0$, which we 
now assume. 

From the definition of $\lambda_\pm$, we see that $\lambda_+\ge \lambda_-$
and hence $A_+\leq A_-$. So the condition $A_\pm >0$ is equivalent to $A_+>0$, 
which is equivalent to $\lambda_+ B^2 <1$, which yields 
\be
J - (\Xi-S\Gamma)B^2 \ > \  B^2\sqrt{(\Xi-S\Gamma)^2 -\Gamma J}\, . 
\ee
As this inequality requires the left-hand side to be positive, we may take the square of both sides to get the following equivalent bound:
\be\label{Jsquared}
J\left[J+ E^2B^2\Gamma - 2\Xi B^2\right] > 0\, .  
\ee
For the static background assumed here, the definition of $J$ in \eqref{J} becomes
\be
J+ E^2B^2\Gamma = 1 + (B^2-E^2)\ell_{PP} +2\sigma EB\,  \ell_{SP}\, , 
\ee
where, as before, $\sigma$ is the sign of $P$. 
Using this, and $J>0$, we find that the bound \eqref{Jsquared} becomes
\be 
1 > E^2\ell_{PP} + B^2\ell_{SS} + 2\sigma EB\, \ell_{SP}\, , 
\ee
which (after multiplication of both sides by $\mathcal{L}_S$) is precisely 
\eqref{Schell2reduced}. We have now established that the strong-field causality condition 
$A_+>0$ (which implies $A_->0$) coincides (in a static background) with the causality condition \eqref{Schell2},  on the assumption that there is a weak-field limit.   

This weak-field limit assumption may be clarified as follows. An implication of the causality conditions $A_\pm >0$ is that $A_+A_->0$. Only the sum and product of $\lambda_\pm$ are needed to compute $A_+A_-$, and the result is 
\be
A_+ A_- = F/G\, , 
\ee
where 
\be
\begin{aligned}
F=&\  1 -\left[E^2 \ell_{PP} + B^2 \ell_{SS} - 2\sigma EB\,  \ell_{SP}\right] \, ,\\
G=&\  1 +  \left[E^2 \ell_{SS} + B^2 \ell_{PP} + 2\sigma EB\,  \ell_{SP}\right]\, . 
\end{aligned}
\ee
Using the causality/convexity conditions in the form \eqref{Schell1'}, 
we may rewrite $G$ as
\be
G= 1+ \left(E\sqrt{\ell_{SS}} - B \sqrt{\ell_{PP}}\right)^2 
+ EB\left( \sqrt{\ell_{SS} \ell_{PP}} +2\sigma \ell_{SP}\right) \, . 
\ee
As the last term in this expression is non-negative as a consequence of $\Gamma \ge 0$, 
we conclude that $G\ge 0$ in any NLED satisfying the convexity/causality conditons of \eqref{Schell1}.  In this context, therefore,  $A_+A_- >0$ is equivalent to $F>0$, but this is \eqref{Schell2reduced}, which is \eqref{Schell2} for our static background. In other words,  although \eqref{Schell2} is implied by $A_- \ge A_+>0$, it is equivalent to $A_+A_- >0$, which allows $A_+\le A_-<0$ and therefore 
 causality violation.

It appears from this result that the conditions \eqref{Schell1} and \eqref{Schell2} are necessary for causality, but not sufficient. However, any path in field space from a weak-field region to one in which both  $A_+$ and $A_-$ are negative, must pass through a point where $A_+A_-\le0$, which would violate the strict equality $F>0$ of \eqref{Schell2reduced} (and, generically, the weaker $F\ge0$). This argument does not apply if there is no weak-field limit but all such cases known to us fall outside the Pleba\'nski class; they do not have a standard Lagrangian density that is a function of $(S,P)$ only.

\section{From Born to Born-Infeld}\label{sec:Kruglov}

We begin our investigation of the consequences of the causality conditions \eqref{Schell1} and (especially) \eqref{Schell2} by considering the one-parameter family of Lagrangian densities 
mentioned in the Introduction: $\mathcal{L}^{(\xi)} =  T- \sqrt{T^2 -2TS -\xi^2 P^2}$. 

The first derivatives with 
respect to $(S,P)$ are 
\be\label{firstD}
\mathcal{L}_S = \frac{T}{T- \mathcal{L}^{(\xi)}} \, , \qquad \mathcal{L}_P = \frac{\xi^2 P}{T- \mathcal{L}^{(\xi)}}\, .
\ee
We remark here that the condition for a Lagrangian density $\mathcal{L}(S,P)$ to define 
an electromagnetic-duality invariant theory is \cite{Bialynicki-Birula:1984daz}
\be\label{PDE}
\mathcal{L}_S^2 - \frac{2S}{P} \mathcal{L}_S \mathcal{L}_P - \mathcal{L}_P^2 =1\, , 
\ee
and this is satisfied by $\mathcal{L}^{(\xi)}$ only for $\xi=1$, the BI case. 

The second derivatives of $\mathcal{L}^{(\xi)}$ are
\be\label{secondD}
\mathcal{L}_{SS} = \frac{T^2}{[T- \mathcal{L}^{(\xi)}]^3}\, , \qquad 
\mathcal{L}_{SP} = \frac{\xi^2TP}{[T- \mathcal{L}^{(\xi)}]^3}\, , \qquad 
\mathcal{L}_{PP} = \frac{\xi^2T(T-2S)}{[T- \mathcal{L}^{(\xi)}]^3}\,  . 
\ee
Using these expressions, one finds that the convexity/causality conditions \eqref{Schell1} are satisfied for any value of $\xi$; for example, $\mathcal{L}_{PP}\ge0$ since reality of $\mathcal{L}_\xi$ requires $T(T-2S)\ge \xi^2 P^2 \ge0$, and 
\be
\mathcal{L}_{SS}\mathcal{L}_{PP} - \mathcal{L}_{SP}^2 = 
\frac{\xi^2T^2}{[T- \mathcal{L}^{(\xi)}]^4} \ge0 \, . 
\ee

Sinilarly, one finds that the causality condition \eqref{Schell2} yields the following inequality\footnote{Equality is possible for $T=0$ but this case falls outside the ``Pleba{\'n}ski class'' considered in \cite{Schellstede:2016zue}.}:
\be
\label{primc}
\left[T-(S+\sqrt{S^2 + P^2})\right] \left[T-\xi^2 \left(S+\sqrt{S^2 + P^2}\right)\right]>0\, . 
\ee
For the original Born theory ($\xi=0$), this reduces to 
$T-S>\sqrt{S^2+P^2}$, which is violated for sufficiently large $P$, as shown in  \cite{Schellstede:2016zue}. In the Born-Infeld case ($\xi =1$) the left-hand side is a perfect square and the inequality is manifestly satisfied.  For all other cases 
($\xi(1-\xi)\ne 1$) the inequality \eqref{primc} is {\sl violated} whenever we can choose $(S,P)$ such that
\be\label{Clims} 
\left\{\begin{array}{ccc} \frac{T}{\xi^2} < S+ \sqrt{S^2+P^2} <T  & & (\xi>1) \\ \\
T< S+ \sqrt{S^2+P^2} < \frac{T}{\xi^2} & & (\xi<1) \, . \end{array}\right. 
\ee
However, the possible values of $(S,P)$ are restricted by the fact that $\mathcal{L}^{(\xi)}$ is real only when 
$(T^2-2TS - \xi^2P^2)\ge0$, which is equivalent to 
\be\label{Lagineq}
S + \sqrt{S^2+\xi^2 P^2} \le T\, . 
\ee
This must be taken into account in drawing conclusions from \eqref{Clims}: 
\begin{itemize}

\item $\xi>1$. In this case it suffices to  consider $P=0$. The causality inequality of \eqref{primc} 
is violated whenever $2S/T \in (\xi^{-2}, 1)$. These values of $S$ are permitted by the restriction 
$2S/T\le 1$ required for reality of  $\mathcal{L}^{(\xi)}$. Also, the violation of causality can occur for zero 
magnetic field ${\bf B}$; in this case it occurs when $|{\bf E}|^2/T \in (\xi^{-2}, 1)$. 

\item $\xi<1$. For $P=0$ a violation of causality now requires $2S>T$, which is incompatible with the 
restriction $2S<T$  required for reality of $\mathcal{L}^{(\xi)}$, so we need only investigate $P\ne0$. It will suffice 
to consider $S=0$, in which case the causality  inequality of \eqref{primc}  is violated whenever $|P|/T \in (1, \xi^{-2})$. 
Reality of $\mathcal{L}^{(\xi)}$ now imposes the restriction  $|P|/T \le \xi^{-1}$, which excludes some values of $|P|/T$ in the interval $(1,\xi^{-2})$ but allows those in the subinterval $(1, \xi^{-1})$. Causality can be violated for these values but now 
a non-zero magnetic field is needed for causality violation (because $P\ne0$).

\end{itemize}
 It follows from this analysis that Born-Infeld ($\xi=1$) is the {\sl only} causal NLED in the one-parameter 
 family defined by $\mathcal{L}^{(\xi)}$. 

This is a rather surprising result because the 
strong-field limit is the causal BB theory for all $\xi\ne0$. This feature can be seen by consideration 
of the following Lagrangian density, 
equivalent to $\mathcal{L}^{(\xi)}$ but involving a pair of auxiliary scalar fields $(u,v)$: 
\be\label{RT}
\mathcal{L}^{(\xi)}_{(RT)} = T- \frac{T}{2}\left\{ v+ \frac{(1+u^2)}{v}\right\} + vS + \xi\,  uP\, .  
\ee
For $\xi=1$ this is the Ro{\v c}ek-Tseytlin (RT) form of $\mathcal{L}_{\rm BI}$ \cite{Rocek:1997hi}; it is notable that 
it generalises to any $\xi$ but not to the version of $\mathcal{L}^{(\xi)}$ for which $\xi^2$ is replaced by a negative
real parameter.  One advantage of this reformulation is that 
it allows us to take the $T\to0$ limit. As $T$ has dimensions of energy density, this is equivalent to a strong-field limit in which the field energy density goes to infinity for fixed $T$. Provided that $\xi\ne0$
we have, defining $u'= \xi u$
\be\label{BB2}
\lim_{T\to0} \mathcal{L}^{(\xi\ne0)}_{(RT)}  = vS + u' P\, , 
\ee
which is the Lagrangian density found in \cite{Bialynicki-Birula:1992rcm} for BB electrodynamics; the scalar fields $(u,v)$ are now Lagrange multipliers for the constraints $S=0$ and $P=0$. However, when $\xi=0$ we have 
\be\label{Born.MRT}
\lim_{T\to0} \mathcal{L}^{(\xi=0)}_{(RT)}  = vS \, , 
\ee
which was shown in \cite{Mezincescu:2023zny} to be the Lagrangian density for the strong-field limit 
of Born's original theory. 

Thus, both the Born theory and the Born-Infeld theory are exceptional cases within the $\xi$-family. All, except the Born theory, have the causal BB as a strong-field limit, but all except Born-Infeld are acausal.  For the remainder of this section, we shall elaborate on this observation, by recovering it from previous results of \cite{Bialynicki-Birula:1984daz, Russo:2022qvz, Mezincescu:2023zny} on wave propagation in constant uniform background fields, and by providing a Hamiltonian perspective, which is simpler than the Lagrangian perspective in two related ways. The Hamiltonian variables are not subject to inequalities analogous to \eqref{Lagineq} and it is possible to take a $T\to0$ limit without having to introduce constraints analogous to those of \eqref{BB2} or \eqref{Born.MRT}.

\subsection{Hamiltonian formulation}

The RT-type Lagrangian density of \eqref{RT} is a good starting point for the passage to the Hamiltonian formulation because it is linear in $(S,P)$. Let us recall here that variation of $\mathcal{L}^{(\xi)}_{(RT)}$ with respect to the auxiliary fields $(u,v)$ yields algebraic field equations that are jointly equivalent to
\be\label{uvL}
u= \frac{\xi P}{T- \mathcal{L}^{(\xi)}}\, , \qquad v= \frac{T}{T-\mathcal{L}^{(\xi)}}\, , 
\ee
and that substitution for $(u,v)$ in $\mathcal{L}^{(\xi)}_{(RT)}$ yields 
$\mathcal{L}^{(\xi)}$. 

To proceed to the Hamiltonian formulation we first define the electric-displacement field ${\bf D}$ (the Legendre dual to ${\bf E}$) by
\be\label{EandD}
{\bf D} := \frac{\partial}{\partial {\bf E}} \left[\mathcal{L}^{(\xi)}_{(RT)}\right] = v{\bf E} + \xi u {\bf B} \quad 
\Rightarrow\quad {\bf E}= v^{-1} \left({\bf D} -\xi u {\bf B}\right)\, . 
\ee
We then define
\be
 \mathcal{H}^\prime_{(\xi)} := {\bf E}\cdot {\bf D} -\mathcal{L}^{(\xi)}_{(RT)} =
\frac{1}{2v}\left\{ |{\bf D} -\xi u{\bf B}|^2 + T(1+u^2)\right\}  + \frac{v}{2} \left(T+ |{\bf B}|^2\right) -T\, ,  
 \ee
 where the prime is a reminder that this ``Hamiltonian density'' is a function of the auxiliary fields $(u,v)$ in 
 addition to $({\bf D},{\bf B})$; their elimination\footnote{By extremisation of $\mathcal{H}'_{(\xi)}$ with respect to $(u,v)$, not by use of \eqref{uvL}.} yields
 \be
 \mathcal{H}_{(\xi)}  = \sqrt{\left(T+|{\bf D}|^2\right)\left(T+|{\bf B}|^2\right) - 
 \xi^2\left(\frac{T+|{\bf B}|^2}{T+ \xi^2 |{\bf B}|^2}\right) \left({\bf D}\cdot{\bf B}\right)^2} -T\, . 
 \ee
 For the BI theory ($\xi=1$) we have the standard result 
\be\label{HamBI}
\mathcal{H}_{\rm BI} = \sqrt{(T+ |{\bf D}|^2)(T+ |{\bf B}|^2) - ({\bf D}\cdot{\bf B})^2} -T\, , 
\ee
and for the Born theory ($\xi=0$) we have
\be\label{HamBorn}
\mathcal{H}_{\rm Born} = \sqrt{(T+ |{\bf D}|^2)(T+ |{\bf B}|^2)} -T\, . 
\ee

Let us reconsider the strong-field ($T\to0$) limit in this Hamiltonian context.  Provided that $\xi$ is non-zero we find that 
\be
\lim_{T\to 0} \mathcal{H}_{(\xi)}  = |{\bf D}\times{\bf B}|  \qquad (\xi\ne0). 
\ee
This is the Hamiltonian density for the conformal BB electrodynamics, originally found this way \cite{Bialynicki-Birula:1984daz} and later interpreted in \cite{Bialynicki-Birula:1992rcm} as a field theory of ``photon dust''. 
In contrast, for $\xi=0$ we have 
\be\label{zeroTB}
\lim_{T\to0} \mathcal{H}_{\rm Born} = |{\bf D}||{\bf B}|\, ,  
\ee
which is a very different conformal field theory. We thus confirm the exceptional nature of the Born theory in this respect.

The process of elimination of the auxiliary fields $(u,v)$ from  $\mathcal{H}_{(\xi)}^\prime$ determines them as functions of $({\bf D},{\bf B})$:
 \be\label{uvH}
 u= \frac{\xi ({\bf D}\cdot {\bf B})}{T+ \xi^2 |{\bf B}|^2} \, , \qquad 
 v= \frac{T+ \mathcal{H}_{(\xi)} }{ T+ |{\bf B}|^2}\, . 
 \ee
Using these relations we find from \eqref{EandD} that
\be\label{E(DB)}
{\bf E} = (T+ \mathcal{H}_{(\xi)})^{-1} \left(\frac{T+ |{\bf B}|^2}{T+\xi^2|{\bf B}|^2}\right) 
\left[(T+ \xi^2|{\bf B}|^2){\bf D} - \xi^2 ({\bf D}\cdot {\bf B}) {\bf B}\right]\, ,  
\ee
from which it follows that 
\be\label{PinH}
P= \left(\frac{T+|{\bf B}|^2}{T+ \xi^2|{\bf B}|^2}\right) 
\left(\frac{T \, {\bf D}\cdot {\bf B}}{T+ \mathcal{H}_{(\xi)}}\right) \, .  
\ee
In addition, comparison of the expressions for $v$ in \eqref{uvL} and \eqref{uvH} 
yields the relation 
\be\label{reciprocal}
(T-\mathcal{L}^{(\xi)})(T+ \mathcal{H}_{(\xi)}) = T(T+ |{\bf B}|^2)\, ,  
\ee
which, combined with \eqref{PinH}, determines $S$ in terms of $({\bf D},{\bf B})$.

For Born's original theory, the relation \eqref{E(DB)} simplifies to
\be
(\sqrt{T+ |{\bf D}|^2}) {\bf E} = (\sqrt{T+|{\bf B}|^2}) {\bf D}\, , 
\ee
which implies that 
\be
|{\bf D}|^2 = \left(\frac{T}{T-2S}\right) |{\bf E}|^2\, .  
\ee
Recalling that $2S \le T$ is required for reality of $\mathcal{L}_{\rm Born}$, we see that the maximum value of $|{\bf E}|$, for any given ${\bf B}$, corresponds to $|{\bf D}|\to \infty$.  There is therefore no restriction on the range of the Hamiltonian field variables $({\bf D},{\bf B})$, and this is true for all NLED in our $\xi$-family, which explains why the strong-field limit is equivalent to a simple $T\to0$ limit in the Hamiltonian formulation.

\subsection{Wave propagation}

We shall now see how our causality results for the Born-BI interpolation family
are recovered from the causality bounds on the birefringent indices for wave 
propagation in a constant uniform electromagnetic background. 
From the formulae of section \ref{sec:causality}  we find that 
\be
\Xi = \frac{\left(1+\xi^2\right)T -2\xi^2S}{2\left(T^2 -2TS+\xi^2 P^2\right)}\, , \qquad 
\Gamma = \frac{\xi^2}{T^2 -2TS+\xi^2 P^2}\, , 
\ee
and 
\be
J= \frac{\left(T-2S\right)\left(T-2\xi^2 S\right)}{T^2 -2TS+\xi^2 P^2}\, . 
\ee
Using these expressions in \eqref{biref.indices} we find that
\be
(\lambda_+,\lambda_- )= \left\{ \begin{array}{cc}
   (\lambda_1,\lambda_2 )  & \xi\leq 1 \ ,\\
   (\lambda_2,\lambda_1 )  & \xi>1 \ ,
\end{array} \right.
\ee
where
\be
\lambda_1= \frac{1}{T-2S} \, , \qquad \lambda_2= \frac{\xi^2}{T-2\xi^2 S}\, .
\ee
Birefringence occurs when these two indices differ, which they do in this case except when $\xi=1$; this is the well-known result 
that BI is a ``zero-birefringence'' NLED. Notice too that $\lambda_2=0$ for $\xi=0$, which implies that $\omega^2=|{\bf k}|^2$
and hence $v_g=1$ for this polarization; this is a feature of any NLED for which $\mathcal{L}_P\equiv 0$. Using these birefringence indices in the formula of \eqref{A} we have
\be\label{ApmEB}
A_1 = \frac{T-E^2}{T+B^2}\, , \qquad A_2 = \frac{T-\xi^2 E^2}{T+ \xi^2 B^2}\,  ,
\ee
where the correspondence with $A_\pm$ is the same as for $\lambda_\pm$
specified  above.

To determine whether either $A_1$ or $A_2$ can be negative one must take into account that the background fields $(E,B)$ are restricted by the condition $T^2-2TS \ge \xi^2P^2$ required for reality of $\mathcal{L}_\xi$. This is what we did in the analysis of \eqref{primc} (which is essentially the condition $A_+ A_->0$) from which we concluded that BI ($\xi=1$) 
is the only causal case. It is instructive to see how the same conclusion is arrived at in Hamiltonian variables. 

For the static background we find from \eqref{E(DB)} that
\be
E = \left(\frac{T+B^2}{T+\xi^2 B^2}\right) \left(\frac{TD}{T+ \mathcal{H}_{(\xi)}}\right) \, , 
\ee
where, in this background, 
\be 
T+\mathcal{H}_{(\xi)}= 
\sqrt{\frac{T(T+B^2)(T+\xi^2B^2 + D^2)}{T+\xi^2B^2}}\, . 
\ee
We thus find that 
\be\label{Apm}
\begin{aligned}
A_1 &=\,  \bar A_1 \left[ 1 - (1-\xi^2) \frac{B^2D^2}{(T+\xi^2B^2)^2}\right]\, , \\
A_2 &=\, \bar A_2 \left[ 1 + (1-\xi^2) \frac{T D^2}{(T+\xi^2B^2)^2}\right]\, , 
\end{aligned}
\ee
where 
\be
\begin{aligned}
    &\bar A_1 = \frac{T(T+\xi^2 B^2)}{(T+B^2)(T+\xi^2B^2 + D^2)}\, ,\\
     &\bar A_2 = \frac{T}{(T+\xi^2B^2 + D^2)}\, .
\end{aligned}
\ee
For use below we give the approximate results for $(A_1,A_2)$ for a strong-field region of field space for which $D^2\gg B^2 \gg T$:
\be\label{probe}
\begin{aligned}
A_1 &= -\frac{(1-\xi^2)}{\xi^2} \left(\frac{T}{B^2}\right)
\left[ 1  + \mathcal{O}\left( \frac{T}{B^2}, \frac{B^2}{D^2}\right)\right]\, , \\
A_2 &= \frac{T}{B^2}\left[\frac{B^2}{D^2} + \frac{(1-\xi^2)}{\xi^4}\frac{T}{B^2} \right] 
\left[ 1  + \mathcal{O}\left( \frac{T}{B^2}, \frac{B^2}{D^2}\right)\right]\, . 
\end{aligned}
\ee
We now consider implications for the various qualitatively distinct values of $\xi$:
\begin{itemize}
\item $\xi=0$. Born's original theory. In this case
\be\label{BornApm}
A_1 = A_+=\frac{T^2 -B^2 D^2}{(T+B^2)(T+D^2)} \, , \qquad A_2=A_- = 1\,.
\ee
We see that $A_\pm\leq 1$, and  $A_->0$, but $A_+<0$ for $|BD|>T$. The Born theory
allows superluminal propagation for sufficiently strong electric field {\sl in the presence of a magnetic field}, in agreement with \cite{Schellstede:2016zue}. 

The $T\to 0$ limit of \eqref{BornApm} yields
\be\label{ApmT0}
A_+=-1\, , \qquad A_- =1 \qquad (T=0),
\ee
but the $T\to 0$ limit of the Born Lagrangian density used to derive \eqref{ApmEB}, from which we 
deduced \eqref{BornApm}, is a zero Lagrangian, so this result for $T=0$ is not obviously justifiable. However, it can be justified by a direct computation of $A_\pm$ within the Hamiltonian 
formulation using the results of \cite{Mezincescu:2023zny}. The result is that the strong-field limit of Born's theory is ``strongly'' acausal, in the sense that one polarisation is always superluminal.  

\item $0<\xi<1$. We see from \eqref{Apm} that $A_+=A_1$ will be negative for sufficiently large $|BD|$, as for Born's original theory but the value of $|BD|$ needed for superluminal propagation increases with $\xi$, becoming infinite at $\xi=1$. 

We also see from \eqref{probe} that although $A_\pm=0$ in the $T\to0$ limit, $A_+ $
is negative as this limit is approached through the strong-field region with $D^2\gg B^2 \gg T$.
This explains how acausality at strong coupling is consistent with 
causality of the strong-coupling limit.  

\item $\xi=1$; i.e. Born-Infeld. In this case 
\be
A_\pm = \bar A_0 = \frac{T}{T+ B^2 + D^2}\, . 
\ee
There is no birefringence, and superluminal propagation is not possible. In the $T\to0$ limit we get $A_\pm =0$ but this result is again not obviously justifiable, and not only for the reason given above in the context of the Born theory. Here there is the additional problem that the $T\to0$ limit of BI is BB, for which there is no static background solution, so our initial assumption of a static background cannot be valid at $T=0$. To understand the $T\to0$ limit it is necessary to start with the dispersion relations for a non-static background, within the Hamiltonian formation, as done in \cite{Bandos:2023yat,Mezincescu:2023zny}. We pass over this here except to say that wave-propagation in BB is lightlike, but only in the direction 
of the (necessarily non-static) background-field momentum density. 

\item $\xi>1$. In this case $A_2=A_+$, which will be negative for sufficiently large $D$. In contrast to the $\xi<1$ cases, including $\xi=0$, superluminal propagation is possible even in backgrounds with zero magnetic field (in agreement with our earlier Lagrangian analysis). 
In particular, the behaviour of $A_\pm$ as the strong-field limit is approached through 
a region with $D^2\gg B^2 \gg T$ can be read off from \eqref{probe}. As for the $0<\xi<1$ case,  we have $A_\pm=0$ at $T=0$ but $A_+$ is negative for any non-zero $T$ when the ratio $TD^2/B^4$ is sufficiently large. 

\end{itemize}
These results confirm our earlier conclusion that the only causal member of 
the $\mathcal{L}^{(\xi)}$ family is Born-Infeld. It is also the only member of the family 
with a Hamiltonian density that is invariant under the $U(1)$ electromagnetic-duality transformation
\be
({\bf D} + i{\bf B}) \to e^{i\theta}({\bf D} + i{\bf B})\, . 
\ee
This fact suggests that we examine some other duality-invariant NLED. Of the few explicitly-known 
examples, the simplest is ModMax and its BI-type duality-invariant extension \cite{Bandos:2020jsw,Bandos:2020hgy}, which we refer to here as ModMaxBorn.


\section{ModMax and ModMaxBorn}

Following \cite{Bandos:2021rqy}, we begin by considering the Lagrangian density
\be\label{ModMax}
\mathcal{L}(S,P) =a \, S +b\sqrt{S^2+P^2}\, ,  
\ee
where $(a,b)$ are arbitrary real constants. The first derivatives of $\mathcal{L}$ are
\be
\mathcal{L}_{S}=a+\frac{bS}{\sqrt{S^2+P^2}}\, , \qquad \mathcal{L}_P = \frac{bP}{\sqrt{S^2+P^2}}\, . 
\ee
In order to have $\mathcal{L}_S>0$ for all $(S,P)$ we need $a>0$ and $b^2<a^2$. 
The second derivatives are 
\be
\mathcal{L}_{SS}=\frac{bP^2}{(S^2+P^2)^{\frac32}}\, , \qquad \mathcal{L}_{SP}=-\frac{bS P}{(S^2+P^2)^{\frac32}}\, , \qquad \mathcal{L}_{PP}=\frac{bS^2}{(S^2+P^2)^{\frac32}}\, . 
\ee
Since $\mathcal{L}_{SS}\mathcal{L}_{PP} = \mathcal{L}_{SP}^2$ (which is a consequence of conformal invariance) the convexity/causality conditions of \eqref{Schell1} require 
$b\ge0$, and the combined constraints on $(a,b)$ become 
\be 
0 \le b < a \, .  
\ee
The general solution to these inequalities may be parametrised as follows: 
\be
a= c \cosh\gamma\, , \qquad b = c\sinh\gamma \, ,
\ee
where $c$ is an arbitrary {\sl positive} constant that determines the overall normalisation of $\mathcal{L}$, and $\gamma$ is a {\sl non-negative} coupling constant.  The choice $c=1$ yields the ModMax Lagrangian density \cite{Bandos:2020jsw}
\be
\mathcal{L}_{\rm MM} = (\cosh\gamma)S + (\sinh\gamma)\sqrt{S^2+P^2} \, . 
\ee

We still have to consider the more complicated causality condition of \eqref{Schell2}. For any conformal theory this reduces to\footnote{This can be proved using relations derived in \cite{Bandos:2021rqy} from the degree-1 homogeneity of $\mathcal{L}(S,P)$.} 
\be
\mathcal{L}_S > (\mathcal{L}_{SS} + \mathcal{L}_{PP}) 
\left(\sqrt{S^2+P^2} +S\right),   
\ee
and for ModMax this is the constraint $a>b$, which is nothing new. In this case
convexity is sufficient for causality. This was also the conclusion of \cite{Bandos:2021rqy} but arrived at by consideration of the birefringence 
indices; from which one finds that 
\be\label{mmlb}
\lambda_-=0,\qquad \lambda_+=\frac{b}{a\sqrt{P^2+S^2}-b S}=\frac{2b}{(a+b)B^2+(a-b)E^2}\, . 
\ee
In the last equality we assumed parallel $({\bf E},{\bf B})$ with magnitudes $(E,B)$, in order to apply the formula of \eqref{A}; for the above birefringence indices this formula yields
\be\label{MM-biref}
A_-=1\, , \qquad A_+ = \frac{a-b}{a+b} = \frac{1-\tanh\gamma}{1+\tanh\gamma}\, .  
\ee
Notice that  $0< A_\pm \le 1$. This tells us that wave propagation in a static homogeneous ModMax background is 
causal. The fact that $A_-=1$ (which follows from $\lambda_-=0$) tells us that one polarisation is lightlike.

We now turn to ModMaxBorn; the Lagrangian density is \cite{Bandos:2020hgy}
\be
\mathcal{L}_{\rm MMB}= T-\sqrt{T^2- 2T\left[(\cosh\gamma) \, S +(\sinh\gamma )\sqrt{S^2+P^2}\right]-P^2}\, . 
\ee
This reduces to $\mathcal{L}_{\rm MM}$ in the weak-field ($T\to\infty$) limit, which is 
causal for $\gamma\ge0$. This is therefore a causality constraint on ModMaxBorn, but is it sufficient for causality? The main task of this section is to prove that it is. 

It is straightforward to show that ModMaxBorn satisfies the convexity/causality conditions of \eqref{Schell1} for $\gamma\ge0$.  We could now proceed to a direct check of whether \eqref{Schell2} is also satisfied, but it is simpler in this case to check strong-field causality via a computation of the birefringence indices; we pass over the straightforward 
but tedious details to give the results. The birefringence indices are
\be
\begin{aligned}
&\lambda_+=\frac{\sqrt{P^2+S^2}+T \sinh (\gamma )}{T \cosh (\gamma )
  \sqrt{P^2+S^2}-S T \sinh (\gamma )-2 S \sqrt{P^2+S^2} }\, , 
\\
&\lambda_- =\frac{\cosh (\gamma ) \sqrt{P^2+S^2}+S \sinh (\gamma )}{T \sqrt{P^2+S^2}-\sinh (\gamma ) \left(P^2+2 S^2\right)-2 S \cosh (\gamma )
   \sqrt{P^2+S^2}}\, . 
\end{aligned}
\ee
In the $T\to\infty$ limit we recover the  birefringence indices \eqref{mmlb} of ModMax. 
For $\gamma=0$ we recover the BI birefringence indices: $\lambda_\pm=1/(T-2S)$.

For a static background field configuration with parallel electric and magnetic fields, the ModMaxBorn birefringence indices simplify to
\be
\begin{aligned}
\lambda_+ &=\ \frac{B^2+E^2+2T\sinh(\gamma)}{B^4-E^4+T(B^2 e^{\gamma}+E^2 e^{-\gamma})}\, , \\
\lambda_- &=\ \frac{B^2+e^{2\gamma} E^2}{(B^4-E^4e^{2\gamma}+T e^{\gamma}(B^2 +E^2 )}\, .
\end{aligned}
\ee
The ModMax results of \eqref{mmlb} are reproduced in the $T\to\infty$ limit, as expected. 
Using this result to compute $A_\pm$, we find that 
\be\label{MMB-biref}
A_-=\frac{Te^\gamma - e^{2\gamma}E^2}{Te^{\gamma}+B^2}\ ,\qquad A_+= e^{-2\gamma}A_-\, . 
\ee
For the same static background we have
\be
\mathcal{L}_{\rm MMB}= T-\sqrt{(T- e^\gamma E^2)(T+e^{-\gamma} B^2)}\, , 
\ee
from which we see that reality requires $e^\gamma E^2\le T$, and hence $A_\pm \ge 0$. 
We also have $A_+<A_- \le1$ provided that $\gamma\ge 0$. We thus conclude that  ModMaxBorn is a 
causal theory because 
\be
0 \le A_\pm \le 1\, . 
\ee
As already mentioned, the possibility of $A_\pm=0$  is not realizable within a standard Lagrangian formulation, 
but we suspect that it could be realized by limits of ModMaxBorn within the Hamiltonian formulation, as discussed
for BI in \cite{Mezincescu:2023zny}.

\section{Causality without Duality}

In order to dispel any idea that causality {\sl requires} duality invariance, 
we now discuss a particular two-parameter family of NLED theories proposed
by Kruglov  \cite{Kruglov:2017mpj} as a generalisation of his earlier
one-parameter family that we analysed in section \ref{sec:causality}. As we shall see, 
this contains a one-parameter subfamily that is causal within a 
parameter range that includes BI. 

The Lagrangian density for the two-parameter family is
\be\label{2param}
\mathcal{L}= \frac{T}{2q}\left(1- \Delta^q\right) \ ,\qquad \Delta\equiv 1 - \frac{2S}{T} - a\, \frac{P^2}{T^2} \, . 
\ee
The weak-field ($T\to\infty$) limit is Maxwell for any $q$. For $q=\tfrac12$ we have the one-parameter family that we analysed in section \ref{sec:Kruglov} (where we replaced $a$ by $\xi^2$). We exclude $q=0$ because this case is essentially the ``logarithmic electrodynamics'' that we consider below in a separate subsection. Notice that 
\be
\mathcal{L}_S = \Delta^{q-1}\, .   
\ee
For $q=1$ the condition $L_S>0$ is trivially satisfied, but then $a=0$ yields the free-field Maxwell theory and the Lagrangian density for $a\ne0$ has the form $\mathcal{L}=F(S)+G(P)$ 
with polynomial $G$; as explained in the Introduction, such cases trivially 
fail the strong-field causality test. Thus, $q=1$ may be excluded.

For $q\ne1$, the $\mathcal{L}_S>0$ condition requires positive $\Delta$, i.e. 
\be
T^2 -2TS -aP^2 >0\, \qquad (q\ne1). 
\ee
To apply the convexity/causality conditions of \eqref{Schell1} we need the following 
quantities:
\be
\begin{aligned}
\ell_{SS} &\ =2(1-q)T^{-1} \Delta^{-1}\ , \\
\ell_{SP} &\ =2a(1-q) PT^{-2} \Delta^{-1}\ ,\\
\ell_{PP} &= a T^{-1}\ \Delta^{-1} \left[\Delta+2a(1-q)\frac{P^2}{T^2}\right]\ , 
\end{aligned}
\ee
which yields
\be
\Gamma \equiv \ell_{SS}\ell_{PP}-\ell_{SP}^2 \, =\, 2a(1-q)\,  T^{-2}\Delta^{-1}\ .
\ee
We see from these results that the convexity/causality conditions \eqref{Schell1} are 
satisfied for $q\ne1$ if and only if 
\be\label{convex}
a\ge0\ ,\qquad  q <1\ .  
\ee

We now need to determine whether additional restrictions on parameters are required by the 
strong-field causality inequality \eqref{Schell2}. Again excluding $q=1$, which we have already dealt with, we find (after dividing by the positive factor $\Delta^{q-2}$) that 
\be\label{tres}
(T-2 a V )\, \left[ 2 (2 q-1)U   (T-2 a V )+T (T-2 V )\right]>0\, ,
\ee
where 
\be
U =\frac12\left(\sqrt{S^2+P^2}- S\right) \, ,\qquad 
V =\frac12\left(\sqrt{S^2+P^2}+ S\right) \, . 
\ee
Notice that $U$ and $V$ are both non-negative,  and that 
\be\label{TDelta}
T^2\Delta = (T-2V )(T+2U ) + 4(1-a) UV  \, .  
\ee
To deduce the implications of \eqref{tres} we must take into account restrictions on the domain of $\mathcal{L}$. For any non-integer $q$, reality of $\mathcal{L}$ requires $\Delta>0$ but   
this is equivalent, for {\sl any} $q<1$,  to the condition $\mathcal{L}_S>1$, which can be expressed as the following upper-bound on $V$:
\be
V   < V_0 \, , \qquad V_0 := \frac{T(T+2U )}{2(T+2aU )}\, . 
\ee

Let us first consider $a\ne1$. As $V <V_0$, we may probe the strong-field region in field-space by choosing
$V =V_0-T \epsilon$, with $0<\epsilon\ll 1$. Expanding the left-hand side of \eqref{tres} in powers of $\epsilon$, 
we arrive at the following version of the strong-field causality inequality, valid to leading order in a power-series expansion in $\epsilon$:
\be\label{probex}
-\frac{4(a-1)^2(1-q)T^4U }{(T+2aU )^4}+O(\epsilon ) >0\, . 
\ee
This inequality is violated for $q<1$.  We thus conclude that all members of the 2-parameter family
with $q<1$ and $a\ne1$ are acausal.  We learn nothing from \eqref{probex} when $a=1$, but in this case the strong-field causality inequality \eqref{tres} simplifies to 
\be\label{tres1}
(T-2V )^2 \left[T+2(2q-1)U \right] >0\,  \qquad (a=1). 
\ee
As $2V<T$ is required for $\Delta>0$ at $a=1$, this inequality is satisfied 
for $q<1$ provided $2q\ge1$. 

We have now shown that all members of the initial 2-parameter NLED family 
defined by the Lagrangian density \eqref{2param} are acausal except for 
the one-parameter subfamily defined by 
\be
\frac12 \leq q< 1\, , \qquad a=1\, . 
\ee
This family contains BI as the $q=\tfrac12$ case. It is perhaps noteworthy that $a=1$ is the value for which $\Delta = -\det(\eta+F/T)$,  which yields the BI Lagrangian density in the form of \eqref{BI.det} for $q=\tfrac12$.  

We conclude by showing that BI is the only member of this one-parameter family of causal NLED theories that is electromagnetic-duality invariant. As mentioned in section \ref{sec:Kruglov},  a Lagrangian density $\mathcal{L}(S,P)$ will define a duality invariant NLED theory only if 
it satisfies the PDE \eqref{PDE}. For independent variables $(U,V)$ this PDE takes the very simple form \cite{Gibbons:1995cv} 
\be
\mathcal{L}_U \mathcal{L}_V =-1 \,, \ \qquad \left(\mathcal{L}_U := \frac{\partial\mathcal{L}}{\partial U}\, , \quad  \mathcal{L}_V := \frac{\partial\mathcal{L}}{\partial V}\right).  
\ee 
A calculation for the Lagrangian density of \eqref{2param} with $a=1$ 
yields
\be
\mathcal{L}_U \mathcal{L}_V =- \Delta^{2q-1}\, , 
\ee
and hence only the $q= \frac12$ case (BI) is duality invariant.  For general $a$, $q$, one obtains
\be
\mathcal{L}_U \mathcal{L}_V =- \left(1+\frac{2aU}{T}\right)\left(1-\frac{2aV}{T}\right)\ \Delta^{2q-2}\, ,  
\ee
which shows that in the two-parameter family only $\{a=0, q=1\}$ (Maxwell) and $\{a=1, q=\frac12\}$ (BI) are duality invariant.

 \subsection{Logarithmic electrodynamics}

A case that is closely related to those considered above has 
\be
\mathcal{L} = - \frac{T}{2} \ln \left[ 1 - \frac{2S}{T} - a\ \frac{P^2}{T^2}\right]\, , 
\label{logt}
\ee
where $a$ is a dimensionless constant. The $a=0$ case dates back to 1995 \cite{Soleng:1995kn}. The $a=1$ case was introduced in \cite{Gaete:2013dta} (and recently applied in \cite{Gaete:2022lkf}) and extended to arbitrary $a$ in \cite{Kruglov:2014iqa}. For any value of $a$ the weak-field limit is Maxwell and all convexity/causality conditions of \eqref{Schell1} are satisfied. However, the strong-field causality condition \eqref{Schell2} is not satisfied for any $a$, as we now show.

The $a=0$ case is a ``Born-type'' theory ($\mathcal{L}_P\equiv 0$) and hence acausal for reasons
already explained in the Introduction. The $a\ne0$ cases require only slightly more analysis. 
Reality of $\mathcal{L}$ imposes a restriction on the allowed values of $(S,P)$, but these allowed values include $S<0$ with $P=0$, for which \eqref{Schell2} reduces to 
\be
2S+T>0\, , 
\ee
which is violated for $2S<-T$, i.e. for sufficiently large magnetic field. We conclude that logarithmic electrodynamics is acausal for any value of the parameter $a$.

 \section{Summary and Outlook}
 
 Causality is an essential requirement for relativistic field theories,  nonlinear electrodynamics (NLED)
 in particular.  For weak fields it is usually a simple matter to ensure causality. Causality violations requiring 
 strong fields are generally associated (in $\hbar =c=1$ units) with some characteristic energy density scale. 
 When this scale comes from higher-derivative terms there are typically additional propagating modes 
 of negative energy; this is a well-known problem.  In the NLED context, it is the reason why the Lagrangian 
 density is restricted to be a function $\mathcal{L}(S,P)$ of the independent Lorentz invariants $(S,P)$ 
 constructed from the electric and magnetic fields, but {\sl not} their derivatives. In the terminology of 
 \cite{Schellstede:2016zue}, this defines the ``Pleba{\'n}ski  class''  of  NLED theories. 
 
 From a Hamiltonian perspective, the restriction to the Pleba{\'n}ski class ensures that the 
 interactions do not change the canonical structure, which ensures that the local degrees of freedom 
 are  the same as the free Maxwell theory.  Thus, interaction terms in the NLED Lagrangian are not ``higher derivative'' even though they introduce (typically) an energy density scale. Born's original 1933 theory and its subsequent Born-Infeld (BI) modification were the first examples; the Born parameter  (called $T$ here) sets the scale.  However, it was not appreciated at that time by Born, or even much later by many others, that strong-field causality violations invisible to weak-field analysis can emerge at the Born scale. 
 
 We suspect that one of the reasons that strong-field causality has rarely been an issue in past work on 
 NLED is that theorists have focused on the Born-Infeld theory, which happens to be causal for both weak and 
 strong fields. Another reason may be that the NLED of principal phenomenological interest is 
 still the Euler-Heisenberg theory, which is derived from QED as an effective field theory. Whatever the reason, it appears to us that most work over the past few decades on applications of new NLED theories, motivated (as was Born's original model) by phenomenological ideas, has been carried out without awareness of the possibility of strong-field causality violation.
 
 Until relatively recently, work on causality in NLED theories focused on particular theories; the first example may be the proof in \cite{Bialynicki-Birula:1984daz} that BI is a causal theory. In that work the group velocity for propagation of plane-wave perturbations of the optical medium provided by a constant electromagnetic background was calculated for BI. The same calculation for a general NLED theory leads to the conclusion that the propagation is causal provided
 that a particular function $A$ of the background fields takes values in the interval $[0,1]$ \cite{Russo:2022qvz}. The structure of this function is such that weak-field causality violation is associated with $A>1$ whereas $A<0$ requires strong fields. Moreover, $A$ is easily found from the birefringence indices, which are known functions of the first and second derivatives of $\mathcal{L}(S,P)$. It is then a relatively small step to express the causality conditions directly in these terms, and we have done this here. The result, however, 
 was first found, via a different method, by Schellstede et al. \cite{Schellstede:2016zue}. 
 
 Specifically, Schellstede et al. found the necessary and sufficient conditions for causality of any NLED in the Pleba{\'n}ski  class, expressed as inequalities involving the first and second derivatives of $\mathcal{L}(S,P)$. In our review of their result (in our notation) we have separated their causality inequalities into two types. The first type are those of \eqref{Schell1}, which are the weak-field causality conditions; they coincide with the conditions for convexity derived in \cite{Bandos:2021rqy} and we have therefore referred to them as the convexity/causality conditions. The second type is the inequality of \eqref{Schell2}, which we have confirmed (using the method sketched above) and 
interpreted as a strong-field causality condition.

Shellstede et al. applied their results to a few cases. In particular, they gave a very simple proof that any NLED theory with a Lagrangian density function $\mathcal{L}(S)$, such as Born's original model, is acausal.  This already rules out as unphysical many NLED theories proposed in the literature; we have mentioned some of them.  A very similar argument can be used for 
Lagrangian density functions of the form $\mathcal{L}=F(S) +G(P)$, and this eliminates a few more NLED theories, but a more detailed analysis is required for models with a more generic Lagrangian density within the Pleba{\'n}ski class. 

One aspect of this analysis is that it is generally necessary to take into account constraints on the domain of the Lagrangian density function $\mathcal{L}(S,P)$ since it must be a real function and this may restrict the allowed values of $(S,P)$. This is nicely illustrated by a NLED model with one free dimensionless parameter, initially proposed by Kruglov, which includes both Born and BI. The Born theory is acausal because, essentially, there is no restriction on $P$; moving in parameter space away from Born towards BI, a restriction on $P$ appears but it is too weak to eliminate the field-space region in which acausality appears until 
BI is reached. Moving beyond BI we now get a restriction on $S$ that is too weak to prevent acausality, which leads to the conclusion that BI is the unique causal member of the family.  

A surprising feature of this Kruglov model, which was our initial motivation for investigating it, is that the strong-field ($T\to0$) conformal limit  is the same for all members of the family except the Born theory, and this limit is the {\sl causal} Bialynicki-Birula electrodynamics. The strong-field conformal limit of Born's theory is a strongly acausal theory, so it is not surprising that Born's theory is itself acausal for sufficiently strong fields.  In contrast, it appears paradoxical that other members of the Born-BI family can be acausal for strong fields but causal in the strong-field limit. The resolution of this paradox becomes clear in the Hamiltonian formulation (which is also much simpler because there are no constraints on the domain of the Hamiltonian density function): although $A\to 0$ in the strong field limit
in all non-BI cases except Born, $A<0$ is always possible before this limit is reached.

We have also analysed a two-parameter extension of the Born-BI family, again proposed by Kruglov \cite{Kruglov:2017mpj}, finding that a one-parameter subset is causal. This subset contains BI as the unique electromagnetic duality-invariant member, thus providing examples of causal theories that are {\sl not} electromagnetic duality invariant.  This is not a surprise because there was never a reason to suppose that electromagnetic duality is required for causality. Nevertheless, it is remarkable that most NLED theories proposed 
in the literature that are not duality invariant are also not causal, whereas the reverse is true for duality invariant theories, as we now explain. 

There are not many known duality-invariant NLED theories; BI is the most well-known.  Recent additions 
are ModMax and its BI-type extension \cite{Bandos:2020jsw,Bandos:2020hgy}; we have proposed the name ModMaxBorn for this NLED family, which has a dimensionless parameter $\gamma$ in addition to a Born scale parameter $T$. In the weak-field limit it reduces to ModMax, which is known to be causal for $\gamma\ge0$ (it reduces to Maxwell at $\gamma=0$) and acausal for 
$\gamma<0$, so we know in advance that ModMaxBorn can be causal only for $\gamma\ge0$. Remarkably, 
this is sufficient for all causality inequalities to be satisfied. ModMaxBorn is causal. In other words, weak-field causality
implies strong-field causality for the entire ModMaxBorn family. As we  show in a separate publication, this is a general feature of self-dual theories
\cite{Russo:2024llm}.

The main purpose of this paper has been to stress the importance of taking into account the possibility of strong-field causality violations in applications of novel NLED theories. In particular, we think that this should be a primary consideration for applications that include gravity. Black Hole singularity theorems can be evaded with simple Born-type NLED theories coupled to gravity \cite{Ayon-Beato:1998hmi},  but the existence of event horizons for black holes depends on the impossibility of acausal propagation.

\section*{Acknowledgements}

PKT has been partially supported by STFC consolidated grant ST/T000694/1. JGR acknowledges financial support from grants 2021-SGR-249 (Generalitat de Catalunya) and MINECO  PID2019-105614GB-C21.


\providecommand{\href}[2]{#2}\begingroup\raggedright

\endgroup


\begin{thebibliography}{10}

\bibitem{Born:1933qff}
M.~Born,
``Modified field equations with a finite radius of the electron,''
Nature \textbf{132} (1933) no.3329, 282.1

\bibitem{Born:1933lls}
M.~Born and L.~Infeld,
``Electromagnetic mass,''
Nature \textbf{132} (1933) no.3347, 970.1

\bibitem{Born:1933pep}
M.~Born and L.~Infeld,
``Foundations of the new field theory,''
Nature \textbf{132} (1933) no.3348, 1004.1

\bibitem{Born:1934gh}
M.~Born and L.~Infeld,
``Foundations of the new field theory,''
Proc. Roy. Soc. Lond. A \textbf{144} (1934) no.852, 425-451


\bibitem{Born:1937drv}
M.~Born,
``Nonlinear theory of the electromagnetic field,''
Ann. Inst. Henri Poincare \textbf{7} (1937) no.4, 155-265



\bibitem{Kuzenko:2000uh}
S.~M.~Kuzenko and S.~Theisen,
``Nonlinear selfduality and supersymmetry,''
Fortsch. Phys. \textbf{49} (2001), 273-309
[arXiv:hep-th/0007231 [hep-th]].

\bibitem{Schrodinger:1935oqa}
E.~Schr\"odinger,
``Contributions to Born's new theory of the electromagnetic field,''
Proc. Roy. Soc. Lond. A \textbf{150} (1935) no.870, 465-477

\bibitem{Heisenberg:1936nmg}
W.~Heisenberg and H.~Euler,
``Consequences of Dirac's theory of positrons,''
Z. Phys. \textbf{98} (1936) no.11-12, 714-732
[arXiv:physics/0605038 [physics]].

\bibitem{Dunne:2012vv}
G.~V.~Dunne,
``The Heisenberg-Euler Effective Action: 75 years on,''
Int. J. Mod. Phys. A \textbf{27} (2012), 1260004
[arXiv:1202.1557 [hep-th]].

\bibitem{Dirac:1960}
P.A.M. Dirac, ``A reformulation of the Born-Infeld Electrodynamics'', 
Proc. Roy. Soc. Lond. A \textbf{257} (1960) no. 1288 32-43

\bibitem{Boillat:1966eyw}
G.~Boillat,
``Vitesses des ondes \'electrodynamiques et lagrangiens exceptionnels,''
Ann. Inst. H. Poincare Phys. Theor. \textbf{5} (1966) no.3, 217-225; 

\bibitem{Boillat:1970gw}
G.~Boillat,
``Nonlinear electrodynamics - Lagrangians and equations of motion,''
J. Math. Phys. \textbf{11} (1970) no.3, 941-951

\bibitem{Plebanski:1970zz}
J.~Pleba{\'n}ski, ``Lectures on non-linear electrodynamics'', (The Niels Bohr Institute and NORDITA, Copenhagen, 1970).

\bibitem{Bialynicki-Birula:1984daz}
I.~Bialynicki-Birula,
``Nonlinear Electrodynamics: Variations on a theme by Born and Infeld'', 
in {\sl Quantum Theory of Particles and Fields}, eds. B. Jancewicz and 
J. Lukierski, (World Scientific, 1983) pp. 31-48. 



\bibitem{Fradkin:1985qd}
E.~S.~Fradkin and A.~A.~Tseytlin,
``Nonlinear Electrodynamics from Quantized Strings,''
Phys. Lett. B \textbf{163} (1985), 123-130

\bibitem{Leigh:1989jq}
R.~G.~Leigh,
``Dirac-Born-Infeld Action from Dirichlet Sigma Model,''
Mod. Phys. Lett. A \textbf{4} (1989), 2767

\bibitem{Tseytlin:1999dj}
A.~A.~Tseytlin,
``Born-Infeld action, supersymmetry and string theory,''
[arXiv:hep-th/9908105 [hep-th]].

\bibitem{Gibbons:2001gy}
G.~W.~Gibbons,
``Aspects of Born-Infeld theory and string/M theory,''
AIP Conf. Proc. \textbf{589} (2001) no.1, 324-350
[arXiv:hep-th/0106059 [hep-th]].

\bibitem{Altshuler:1990}
B.L.~Altshuler, ``An alternative way to inflation and the possibility of anti-inflation'', Class. Quant. Grav. 7, 189 (1990).



\bibitem{Ayon-Beato:1998hmi}
E.~Ayon-Beato and A.~Garcia,
``Regular black hole in general relativity coupled to nonlinear electrodynamics,''
Phys. Rev. Lett. \textbf{80} (1998), 5056-5059
[arXiv:gr-qc/9911046 [gr-qc]].

\bibitem{Bronnikov:2000vy}
K.~A.~Bronnikov,
``Regular magnetic black holes and monopoles from nonlinear electrodynamics,''
Phys. Rev. D \textbf{63} (2001), 044005
[arXiv:gr-qc/0006014 [gr-qc]].

\bibitem{Maeda:2008ha}
H.~Maeda, M.~Hassaine and C.~Martinez,
``Lovelock black holes with a nonlinear Maxwell field,''
Phys. Rev. D \textbf{79} (2009), 044012
[arXiv:0812.2038 [gr-qc]].


\bibitem{Uniyal:2022vdu}
A.~Uniyal, R.~C.~Pantig and A.~\"Ovg\"un,
``Probing a non-linear electrodynamics black hole with thin accretion disk, shadow, and deflection angle with M87* and Sgr A* from EHT,''
Phys. Dark Univ. \textbf{40} (2023), 101178
[arXiv:2205.11072 [gr-qc]].

\bibitem{Tomizawa:2023vir}
S.~Tomizawa and R.~Suzuki,
``Causality of photon propagation under dominant energy condition in nonlinear electrodynamics,''
Phys. Rev. D \textbf{108} (2023) no.12, 124072
[arXiv:2309.10535 [gr-qc]].


\bibitem{Gaete:2021ytm}
P.~Gaete and J.~A.~Helay\"el-Neto,
``Remarks on inverse electrodynamics,''
Eur. Phys. J. C \textbf{81} (2021) no.10, 899
[arXiv:2108.07929 [hep-ph]].

\bibitem{Lai:2022tjr}
C.~Lai and Q.~Pan,
`Complexity for holographic superconductors with the nonlinear electrodynamics,''
Nucl. Phys. B \textbf{974} (2022), 115615

\bibitem{Sorokin:2021tge}
D.~P.~Sorokin,
``Introductory Notes on Non-linear Electrodynamics and its Applications,''
Fortsch. Phys. \textbf{70} (2022) no.7-8, 2200092
[arXiv:2112.12118 [hep-th]].



\bibitem{Mezincescu:2023zny}
L.~Mezincescu, J.~G.~Russo and P.~K.~Townsend,
``Hamiltonian birefringence and Born-Infeld limits,''
[arXiv:2311.04278 [hep-th]].

\bibitem{Schellstede:2016zue}
G.~O.~Schellstede, V.~Perlick and C.~L\"ammerzahl,
``On causality in nonlinear vacuum electrodynamics of the Pleba\'nski class,''
Annalen Phys. \textbf{528} (2016) no.9-10, 738-749
[arXiv:1604.02545 [gr-qc]].


\bibitem{Bandos:2021rqy}
I.~Bandos, K.~Lechner, D.~Sorokin and P.~K.~Townsend,
``ModMax meets Susy,''
JHEP \textbf{10} (2021), 031
[arXiv:2106.07547 [hep-th]].

\bibitem{Russo:2022qvz}
J.~G.~Russo and P.~K.~Townsend,
``Nonlinear electrodynamics without birefringence,''
JHEP \textbf{01} (2023), 039
[arXiv:2211.10689 [hep-th]].

\bibitem{Novello:1998ws}
M.~Novello, J.~M.~Salim, V.~A.~De Lorenci and R.~Klippert,
``Effective Lagrangian for electrodynamics and avoidance of the singular origin of the universe,''
[arXiv:gr-qc/9806076 [gr-qc]].


\bibitem{Gibbons:2001sx}
G.~W.~Gibbons and C.~A.~R.~Herdeiro,
``The Melvin universe in Born-Infeld theory and other theories of nonlinear electrodynamics,''
Class. Quant. Grav. \textbf{18} (2001), 1677-1690
[arXiv:hep-th/0101229 [hep-th]].

\bibitem{Kruglov:2014iwa}
S.~I.~Kruglov,
``Nonlinear arcsin-electrodynamics,''
Annalen Phys. \textbf{527} (2015), 397-401
[arXiv:1410.7633 [physics.gen-ph]].


\bibitem{Kruglov:2009he}
S.~I.~Kruglov,
``On generalized Born-Infeld electrodynamics,''
J. Phys. A \textbf{43} (2010), 375402
[arXiv:0909.1032 [hep-th]].

\bibitem{Bandos:2020jsw}
I.~Bandos, K.~Lechner, D.~Sorokin and P.~K.~Townsend,
``A non-linear duality-invariant conformal extension of Maxwell's equations,''
Phys. Rev. D \textbf{102} (2020), 121703
[arXiv:2007.09092 [hep-th]].

\bibitem{Fox:1970cu}
R.~Fox, C.~G.~Kuper and S.~G.~Lipson,
`Faster-than-light group velocities and causality violation,''
Proc. Roy. Soc. Lond. A \textbf{316} (1970), 515-524

\bibitem{Bandos:2020hgy}
I.~Bandos, K.~Lechner, D.~Sorokin and P.~K.~Townsend,
``On p-form gauge theories and their conformal limits,''
JHEP \textbf{03} (2021), 022
[arXiv:2012.09286 [hep-th]].

\bibitem{Rocek:1997hi}
M.~Rocek and A.~A.~Tseytlin,
``Partial breaking of global D = 4 supersymmetry, constrained superfields, and three-brane actions,''
Phys. Rev. D \textbf{59} (1999), 106001
[arXiv:hep-th/9811232 [hep-th]].

\bibitem{Bialynicki-Birula:1992rcm}
I.~Bialynicki-Birula,
``Field theory of photon dust,''
Acta Phys. Polon. B \textbf{23} (1992), 553-559.

\bibitem{Bandos:2023yat}
I.~Bandos, K.~Lechner, D.~Sorokin and P.~K.~Townsend,
``Trirefringence and the M5-brane,''
JHEP \textbf{06} (2023), 171
[arXiv:2303.11485 [hep-th]].

\bibitem{Kruglov:2017mpj}
S.~I.~Kruglov,
``Born\textendash{}Infeld-type electrodynamics and magnetic black holes,''
Annals Phys. \textbf{383} (2017), 550-559
[arXiv:1707.04495 [gr-qc]].

\bibitem{Gibbons:1995cv}
G.~W.~Gibbons and D.~A.~Rasheed,
``Electric - magnetic duality rotations in nonlinear electrodynamics,''
Nucl. Phys. B \textbf{454} (1995), 185-206
[arXiv:hep-th/9506035 [hep-th]].

\bibitem{Soleng:1995kn}
H.~H.~Soleng,
``Charged black points in general relativity coupled to the logarithmic U(1) gauge theory,''
Phys. Rev. D \textbf{52} (1995), 6178
[arXiv:hep-th/9509033 [hep-th]].

\bibitem{Gaete:2013dta}
P.~Gaete and J.~Helay\"el-Neto,
``Finite Field-Energy and Interparticle Potential in Logarithmic Electrodynamics,''
Eur. Phys. J. C \textbf{74} (2014) no.3, 2816
[arXiv:1312.5157 [hep-th]].

\bibitem{Gaete:2022lkf}
P.~Gaete and J.~A.~Helay\"el-Neto,
``Vacuum material properties and Cherenkov radiation in logarithmic electrodynamics,''
Eur. Phys. J. C \textbf{83} (2023) no.2, 128
[arXiv:2205.03252 [hep-ph]].

\bibitem{Kruglov:2014iqa}
S.~I.~Kruglov,
``On Generalized Logarithmic Electrodynamics,''
Eur. Phys. J. C \textbf{75} (2015) no.2, 88
[arXiv:1411.7741 [hep-th]].





\bibitem{Russo:2024llm}
J.~G.~Russo and P.~K.~Townsend,
``Causal Self-Dual Electrodynamics,''
Phys. Rev. D \textbf{109} (2024), 105023
[arXiv:2401.06707 [hep-th]]. 













\end{thebibliography}

\end{document}